\documentstyle[12pt,graphicx,caption2]{article}

\begin{document}

\title{Comment to "On the field-theoretical approach to the neutron-antineutron
oscillations in nuclei"}
\author{V.I.Nazaruk\\
Institute for Nuclear Research of RAS, 60th October\\
Anniversary Prospect 7a, 117312 Moscow, Russia.}

\date{}
\maketitle
\bigskip

\begin{abstract}
We briefly outline the present state of the $n\bar{n}$ transition problem
and analyse the recent manuscript of the Vladimir Kopeliovich (arXiv: 0912.5065).
\end{abstract}

\vspace{1cm}

\setcounter{equation}{0}

First of all we briefly sum up the present state of the investigations of this problem. 

In the standard calculations of $ab$ oscillations in the medium [1-3] the interaction of 
particles $a$ and $b$ with the matter is described by the potentials $U_{a,b}$ (potential 
model). ${\rm Im}U_b$ is responsible for loss of $b$-particle intensity. In particular, 
this model is used for the $n\bar{n}$ transitions in a medium [4-10] followed by annihilation:
\begin{equation}
n\rightarrow \bar{n}\rightarrow M,
\end{equation}
here $M$ are the annihilation mesons.

In [9] it was shown that one-particle model mentioned above does not describe the
total $ab$ (neutron-antineutron) transition probability as well as the channel corresponding 
to absorption of the $b$-particle (antineutron). The effect of final state absorption
(annihilation) acts in the opposite (wrong) direction, which tends to the additional
suppression of the $n\bar{n}$ transition. The $S$-matrix should be unitary.

In [11] we have proposed the model based on the diagram technique which does not contain
the non-hermitian operators. Subsequently, this calculation was repeated in [12]. However, in 
[13] it was shown that this model is wrong: the neutron line entering into the $n\bar{n}$ 
transition vertex should be the {\em wave function}, but not the propagator, as in the model 
based on the diagram technique. For the problem under study this fact is crucial. It leads 
to the cardinal error for the process in nuclei. The $n\bar{n}$ transitions in the medium and 
vacuum are not described at all. If the neutron binding energy goes to zero, the result
diverges (see Eqs. (18) and (19) of Ref. [11] or Eqs. (15) and (17) of Ref. [12]). So we 
abandoned this model [13]. 

In [14] the model which is free of drawbacks given above has been proposed. However, the 
consideration was schematic since our concern was only with the role of the final state 
absorption in principle. For this reason we continue consideration [15] of above-mentioned 
model as well as the model with bare propagator  (in the notations of [15] the models {\bf b}
and {\bf a}, respectively.)

In the recent e-print [16] the previous calculations [11,12] have been repeated. It is easy
to verify that there is nothing new compared with [11,12], with the exception of Sect. 5.
However, the main statement of this section is completely wrong. The author writes: "If
the infrared divergence discussed in [10,13] takes place for the process of $n\bar{n}$ 
transitions in nucleus, it should take place also for the nucleus form-factor at zero
momentum transfer", which is absolutely wrong.

In the model under consideration [11,12,16] the $n\bar{n}$ transition takes place in the loop.
For the propagators in the loop the infrared divergence (for $n\bar{n}$ transition,
nucleus form-factor, and so on) cannot be in principle. In odder to obtain the infrared 
divergence, the neutron line entering into the $n\bar{n}$ transition vertex should be the 
wave function, but not the propagator, as in the model based on the diagram technique.

Thus, on the one hand, in the model based on the diagram technique the infrared divergence 
cannot be in principle for any process including the $n\bar{n}$ transition, and on the other 
hand this model is wrong. The model containing the vertex of virtual decay $A\rightarrow
n+(A-1)$ is unsuitable for the problem under study [13]. In the correct model the neutron 
line entering into the $n\bar{n}$ transition vertex should be the wave function. In this
case we inevitably get the infrared singularity for the plane wave or wave function of bound
state.

Sect. 3 which pretends to the common character of consideration in our opinion has no need of
comments. In particular, in the end of this section we read: "Technical reason for the strange
result obtained in [10,13] is the wrong interpretation of the second order pole structure of
any amplitude containing $n\bar{n}$ transitions." In reality, the technical reason is that
we simply use the fundamentally different process model.

\end{document}